\documentstyle[preprint,prd,aps]{revtex}
\begin{document}
\def\overlay#1#2{\setbox0=\hbox{#1}\setbox1=\hbox to \wd0{\hss #2\hss}#1%
\hskip -2\wd0\copy1}
\author{Yeu-Chung Lin}
\address{
Department of Physics, National Central University, Chung-Li, Taiwan 32054}
\title{
Prediction of the anomalous magnetic moment
of nucleon from the nucleon anomaly}
\date{\today}
\maketitle

\begin{abstract}
We construct the effective anomaly lagrangian involving nucleons and photons by
using current-current coupling method. The contribution of this
lagrangian to the anomalous magnetic moment of nucleon is purely isovector. The
anomalous magnetic moment of proton, $\kappa_P$, can be calculated from the
this lagrangian and it is found
to be $\kappa_P^{Theor.} = 1.77$, which is in excellent agreement with the
experimental
value  $\kappa_P^{Exp.} = 1.79$. While the case of neutron, $\kappa_N^{Theor.}
= -2.58$ as compared to $\kappa_N^{Exp.} =-1.91$, is less satisfactory, but the
sign is correct.
\end{abstract}

\newpage
\narrowtext

\noindent
{\em Introduction}
The Skyrme model\cite{skyr61} reveals the underlying $QCD$ physics of baryon
in the large
$N_c$ limit by considering baryon as a soliton made of mesons. The Skyrme
lagrangian, inheriting it's form from the meson lagrangian, contains two parts,
the normal
parity part, which is basically the kinetic energy term, and the abnormal
parity part, which is known as $W.Z.W.$\cite{wess71,witt83} terms. The
electromagnetic interaction can be incorporated into the Skyrme lagrangian by
the
same way as what the meson lagrangian does. It is interesting to observe that
both the $\gamma PP$ and $\gamma\gamma PP$ vertices appear in both normal and
abnormal parity parts of Skyrme lagrangian, as contrary to what we learned from
the meson lagrangian that amplitudes never receive mixed contribution from
different parity sectors. The
reason for this difference is quite simple. The notion of $G$-parity failes to
work for baryon, and hence the distinction between normal and abnormal parity
is meaningless in the baryon case. Nevertheless, the existence of $\gamma PP$
and $\gamma\gamma PP$ vertices in the $W.Z.W.$ terms indicates that these two
vertices could receive contribution from anomaly. It is the main goal of the
present work to construct an effective lagrangian to describe the anomaly
effect which contributes to $\gamma PP$ and $\gamma\gamma PP$ vertices.
And then use the lagrangian so constructed to predict the anomalous magnetic
moment of proton.
\vspace{1cm}

\noindent
{\em Construction of Effective Lagrangian}
A heuristic way for the construction is to consider the formation of
the triangle anomaly.
The structure of triangle anomaly is determined by the triangle loop and
the parity of the $AVV$ external currents. It is insensitive to the actual
contents of the external currents. Replacement of one of the photon by $Z$
(which also contains a vector part)
in $\pi^0\rightarrow\gamma\gamma$, for example, will not spoil the fact
of being an anomaly.

Inspired by the above observation, we can use the
$\pi^0\rightarrow\gamma\gamma$ anomaly as the building block of the
construction of effective lagrangian involving nucleons.
Namely, we can cast the amplitude of $\pi^0\rightarrow\gamma\gamma$ anomaly
into a current-current coupling form
\begin{eqnarray}
A(\pi^0\rightarrow\gamma\gamma) & = & -f_{\pi}\partial_{\mu}\pi^0\cdot
{e^2 \over 4\pi^2 f_{\pi}^2}
\epsilon^{\alpha\beta\mu\nu} A_{\nu}\partial_{\alpha}A_{\beta},
\end{eqnarray}
and then replace the pion axial current by the nucleon axial current. The
result of this substitution can be used to describe the nucleon anomaly.
The relative normalization of the neutral axial currents of pion and nucleon is
determined by the axial current derived from the $\pi$-N lagrangian. The
neutral axial current is given by
\begin{eqnarray}
J^5_{\mu}  & = & -f_{\pi}\partial_{\mu}\pi^0 +
\bar{\psi}\gamma_{\mu}\gamma_5 {\tau^3\over 2}\psi.
\end{eqnarray}
Replacing the pion current in $eq.$ 1 by the nucleon current in $eq.$ 2, we
derive
a piece in the effective lagrangian which describes the $\gamma\gamma PP$
interaction
\begin{eqnarray}
L_{PP\gamma\gamma} & = & {e^2 \over 4\pi^2 f_{\pi}^2}
\epsilon^{\alpha\beta\mu\nu}\bar{\psi}\gamma_{\mu}\gamma_5
{\tau^3\over 2}\psi A_{\nu}\partial_{\alpha}A_{\beta}.
\end{eqnarray}
This piece of lagrangian is obvious not gauge invariant by itself. By trial
and error, we can add another piece to make it gauge invariant
\begin{eqnarray}
L & = & {e^2 \over 4\pi^2 f_{\pi}^2}
\epsilon_{\alpha\beta\mu\nu}\bar{\psi}\gamma_{\mu}\gamma_5
{\tau^3\over 2}\psi A_{\nu}\partial_{\alpha}A_{\beta}\nonumber\\
            &  & +{e \over i4\pi^2
f_{\pi}^2}\epsilon_{\alpha\beta\mu\nu}\bar{\psi}\gamma_{\mu}\gamma_5
{\tau^3\over 2}\partial_{\nu}\psi \partial_{\alpha}A_{\beta} .
\end{eqnarray}
The effective lagrangian so constructed describes the anomalous interactions of
nucleons and photons.

It is natural to ask whether the single photon piece in $eq.$ 4 is a double
counting of $\gamma PP$ vertex, since the vertex can also arise from the
minimum coupling
of Dirac term. The answer is of course no, as we have learned from the Skyrme
lagrangian that such vertex could exit in both places. An explicit way to
show that the double counting does not occur is to decompose the single
photon piece into two part, a part proportional to $\gamma_{\mu}$ and a part
proportional to $\sigma_{\mu\nu}$, and then check whether the $\gamma_{\mu}$
piece contributes to $F_1 (0)$ in Rosenbluth formular. The normalization
of proton charge requires $F_1 (0)$ remains to be one. We found that indeed the
single photon piece does not contribute to $F_1 (0)$.
\vspace{1cm}

\noindent
{\em Anomalous Magnetic Moment}
The anomalous magnetic moment of nucleon can not arise from the minimum
coupling of Dirac term. Traditionally, the term which gives the
the anomalous magnetic moment of nucleon is written in a gauge invariant form
\begin{eqnarray}
L & = & -{\kappa\mu_B \over 2}
\bar{\psi}\sigma_{\mu\nu}F^{\mu\nu}\psi.
\end{eqnarray}
The anomalous magnetic moment $\kappa$ is left as a free parameter which is to
be determined. Many predictions of the anomalous magnetic moment are
actually symmetry statement
in nature, namely, they are the just simultaneous fittings of the anomalous
magnetic
moment of baryon octet by using $SU(3)$ relation. So the success of predictions
is rather
an evidence of hadron symmetry than the understanding of hadron structure.
It is the purpose of the present study to {\it predict} the anomalous magnetic
moment of nucleon by relating it to the nucleon anomaly.

Since we start with a very different approach of constructing lagrangian, it is
not
transparent to see how can we compare our result with the form
in $eq.$ 5. An essential definition of anomalous magnetic
moment can be defined through the magnetic structure function of nucleon
probed in the differential cross section of electron-nucleon elastic
scattering. Using the lagrangian in $eq.$ 4,
it is a simple exercise to calculate the differential cross section
of $e^- P\rightarrow e^- P$
\begin{eqnarray}
{d\sigma \over d\Omega} & = & \left( {d\sigma \over d\Omega} \right)_{Mott}
\left\{ \left[ 1-{q^2\over 4M^2} \left( {M^2\over 4\pi^2 f_{\pi}^2} \right)
\right]
\right. \nonumber\\
&  & + \left. -{q^2\over 4M^2} \cdot 2 \left[ 1 + \left( {M^2\over
4\pi^2 f_{\pi}^2} \right)^2 \right] \tan^2{\theta \over 2} \right\},
\end{eqnarray}
where $\left( {d\sigma \over d\Omega} \right)_{Mott} $ is what one would get
for
a spinless point charge. The magnetic structure function of proton can be read
off from the second line in the above equation
\begin{eqnarray}
{G_M^P}  & = &  \left[ 1 + \left({M^2\over
4\pi^2 f_{\pi}^2}\right)^2\right]^{1/2}.
\label{eq8}
\end{eqnarray}
It is related to the anomalous magnetic moment of proton
${G_M^P}^2=(1+\kappa_P)^2$. Numerically we find $\kappa_P=1.77$ which is
in excellent agreement with the experimental data $\kappa_P=1.79$.
The magnetic structure function of neutron can be derived in the same way and
it is
given by
\begin{eqnarray}
{G_M^N}  & = &  - {M^2\over
4\pi^2 f_{\pi}^2}.
\end{eqnarray}
The numerical value $\kappa_N=-2.58$ deviates from the
experimental data $\kappa_N=-1.91$ by $35\%$,
which is less satisfactory as compared with the success of the proton case.
Nevertheless, the sign is correct. It is the isovector nature of the
axial current coupling leads to the change of sign for the neutron case. No
isoscarlar contribution is found from the lagrangian.

We want to
stress that the above results are true predictions of hadron structure,
since the sole input
of this prediction is the relative normalization between pion and nucleon axial
currents. This is the main result of the present paper.
\vspace{1cm}

\noindent
{\em Discussion and Conclusion}
The stratedge of constructing an effective lagrangian is to include all the
possible terms allowed by the required symmetries. The conventional form $eq.$
5
is itself a gauge invariant piece and in principle should be included in the
lagrangian at the first sight. But it is redundant to do so as we will see in
the following analysis. Using the identity
\begin{eqnarray}
\epsilon^{\mu\nu\rho\sigma}\gamma_{5}\gamma_{\sigma} & = &
-i\gamma^{\mu}\gamma^{\nu}\gamma^{\rho} + i g^{\mu\nu}\gamma^{\rho}
-i g^{\mu\rho}\gamma^{\nu} + ig^{\nu\rho}\gamma^{\mu},
\end{eqnarray}
the single photon piece in the lagrangian can be decomposed
into two pieces
\begin{eqnarray}
&& {eM \over i4\pi^2 f_{\pi}^2}\bar{\psi}\sigma_{\alpha\beta}{\tau^3\over 2}
F_{\alpha\beta}\psi\nonumber\\
&& + {eM \over 4\pi^2 f_{\pi}^2}\bar{\psi}{\tau^3\over 2}[\partial^{\alpha}\psi
\partial_{\alpha}\!\not\!\! A - \partial^{\beta}\psi\not\!\partial A_{\beta}].
\end{eqnarray}
The first term is gauge invariant by itself and is of the conventional form.
So the conventional form has been included implicitly in the lagrangian
already. However, it is not proper to identify the coefficient
${eM \over i4\pi^2 f_{\pi}^2}$ as $-\kappa \mu_B$, because the second
term also contributes to the magnetic structure function. Due to the
nonrenormalization theorem of anomaly, we conjecture that the coefficients of
the lagrangian is not renormalized by quantum correction.

The lagrangian shares various common results with the Skyrme
model\cite{braa86}. The leading
contribution to the anomalous magnetic moment is isovector in nature and this
leads to the correct sign of the neutron anomalous magnetic moment. However,
the W.Z.W. lagrangian in the Skyrme model only gives arise to the
isoscalar part of the anomalous magnetic moment and it is numerically small
compared with the isovector part arising from the kinetic energy term. The
source of the difference remains an issue to be further studied.

Replacing the pion current by nucleon current in triangle anomaly, we derived a
minimum gauge invariant set of effective lagrangian which describes the
anomalous interaction of photons and nucleons. The prediction for the anomalous
magnetic moment of proton is in excellent agreement with the experimental data.
The construction of the effective lagrangian
provides a new way of understanding the hadronic structure of
nucleon.

\section*{Acknowledgments}
This research is supported in part by the National Science Council in Taiwan
under Contract No. NSC84-2112-M008-013 and NSC84-2732-M008-001. The author
likes to thank C.\ Y.\ Cheung and S.\ C.\ Lee for useful conversation.

\end{document}